\documentclass{easychair}
\usepackage{amsmath}
\usepackage{amssymb}
\usepackage{amsthm}
\usepackage[nocompress]{cite}
\usepackage{colortbl}
\usepackage{graphicx}
\usepackage{hyperref}
\usepackage{listings}
\usepackage{multicol}
\usepackage{xcolor}
\usepackage{xspace}
\usepackage{our_definitions}

\allowdisplaybreaks

\makeatletter
\newcommand{\leqnomode}{\tagsleft@true\let\veqno\@@leqno}
\newcommand{\reqnomode}{\tagsleft@false\let\veqno\@@eqno}
\makeatother

\newcommand{\PH}[1]{#1}

\def\orcidID#1{\href{http://orcid.org/#1}{\raisebox{-1.25pt}{\includegraphics{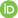}}}}

\theoremstyle{definition}
\newtheorem{example}{Example}

\begin{document}
\title{Overapproximation of Non-Linear Integer Arithmetic for Smart Contract Verification}

\titlerunning{Overapproximation of NIA for Smart Contract Verification}
%
\author{
Petra Hozzov\'a\inst{1}\orcidID{0000-0003-0845-5811} \and
Jaroslav Bend\'ik\inst{2} \and
Alexander Nutz\inst{2}\orcidID{0000-0003-2080-2732} \and
Yoav Rodeh\inst{2}}
\authorrunning{Hozzov\'a et al.}
%
\institute{TU Wien \and Certora}
\maketitle              
\begin{abstract}
%
%
%
The need to solve non-linear arithmetic constraints presents a major obstacle to the automatic verification of smart contracts.
In this case study we focus on the two overapproximation techniques used by the industry verification tool Certora Prover: overapproximation of non-linear integer arithmetic using linear integer arithmetic and using non-linear real arithmetic.
We compare the performance of contemporary SMT solvers on verification conditions produced by the Certora Prover using these two approximations against the natural non-linear integer arithmetic encoding.
Our evaluation shows that the use of the overapproximation methods leads to solving a significant number of new problems.

%
%
\end{abstract}

\section{Introduction}\label{sec:intro}



Smart contracts~\cite{szabo1994} are an ideal target for automated program verification in many ways: Their source code is often relatively small. Due to high execution costs (also known as gas consumption), they contain few loops or recursive methods. Finally, since the underlying applications are often financial in nature, there is a large incentive to detect bugs early, and hacks that lead to a loss of money in the tens of millions of US dollars are not rare.
However, a major pain point is that the verification conditions of smart contracts often contain non-linear arithmetic, which makes them undecidable in general and sometimes intractable in practice.

In this case study, we evaluate two methods to prove verification conditions stemming from smart contract verification that were previously intractable due to their use of non-linear integer arithmetic (NIA).
The methods are based on the observation that the integer inequalities that occur in the smart contract verification conditions can often be effectively reasoned about at lower precision than the one demanded by the standard NIA semantics, thus allowing for a precision-complexity trade-off.
We evaluate two techniques to overapproximate a NIA formula, one using linear integer arithmetic (LIA) and one using nonlinear real arithmetic (NRA).

The motivation leading to this case study was to augment the portfolio of methods that the Certora Prover, a leading industrial tool for automatic verification of smart contracts, uses for verifying properties of Ethereum~\cite{wood2014ethereum} smart contracts containing NIA.
To verify a smart contract property, the Certora Prover builds a logical formula, called a \emph{verification condition (VC)}, such that the VC is satisfiable iff the specification can be violated by the contract.
Moreover, every model of the VC corresponds to a particular counter-example, i.e., an execution of the smart contract violating the specification.
To check the VC satisfiability, the Certora Prover converts the VC to an SMT formula and passes it to an SMT solver, such as \cvc5~\cite{cvc5} or \zsol3~\cite{Z3}.

The smart contract VC is encoded in a fragment of first-order logic with theories, uninterpreted sorts, and uninterpreted function symbols.
Depending on the smart contract, specification, and tool configuration, the VC may contain quantifiers, datatypes, 
LIA or NIA, and bitwise operations.
This study focuses on the cases with quantifier-free VC using NIA.
Since satisfiability of formulas with NIA is undecidable~\cite{matiyasevich1993} it is not too surprising that SMT solvers often struggle with proving such VCs. 
Our evaluation shows this difficulty, as well as a significant number of benchmarks being newly solved using the two overapproximation techniques.

The idea to overapproximate the underlying non-linearity was developed in the Certora Prover already from its conception with a simple LIA overapproximation, which was then gradually refined by adding axioms specialized for the domain of smart contracts. 
Motivated by empirical success of the LIA-based method, we recently implemented another overapproximation based on NRA.
Both 
methods are designed for proving the VC unsatisfiability, not finding counter-examples: overapproximation unsatisfiability implies unsatisfiability of the original VC, but not vice versa.
Interestingly, contemporary SMT solvers also internally apply various overapproximation techniques while solving NIA instances~\cite{cimatti2018,kremer2016generalised,jovanovic2017solving}.
As our results show, these techniques seem to be largely complementary to the overapproximations we investigate in this paper.

\paragraph{Contributions.}
The main focus of this paper is on experimental evaluation of two SMT-based
overapproximation techniques for smart contract verification, employed by the Certora Prover.
Namely, we compare the performance of contemporary SMT solvers on the natural NIA encoding against the LIA and NRA overapproximations.
Using the two overapproximations, we are able to verify 39\% new examples in our benchmark suite, which could not be verified using the NIA encoding.
These benchmarks are real-world benchmarks: 
they stem from actual customer contracts that were under verification by Certora, and are made available online.
Secondary contributions of this paper are (brief) descriptions of the LIA and NRA encodings.

\paragraph{Outline.}
Section~\ref{sec:prelim} lays out the preliminaries and introduces an illustrative example.
Section~\ref{sec:nra} presents the 
overapproximation methods and demonstrates one of them on our example.
Section~\ref{sec:experiments} reports on the experiment setup and results.
Section~\ref{sec:related} briefly reviews related work.

\section{Preliminaries and Example}\label{sec:prelim}

We assume familiarity with standard first-order logic with equality and the theories of integer and real arithmetic (for details see~\cite{gallier2015logic,SMTLIB-standard}).
We use the theories of linear integer arithmetic, and non-linear arithmetic for integers and reals in accordance with SMT-LIB definitions~\cite{SMTLIB}, denoted by LIA, NIA and NRA, respectively.
These theories define the 
predicates less, less equal, greater, greater equal, and the functions of addition, subtraction, (only NIA and NRA) multiplication, division, (only NIA) absolute value\footnote{In SMT-LIB, only NIA defines the absolute value function. However, since $\absi$ does not occur in the VCs produced by the Certora Prover, and since its counterpart for LIA and NRA could be straightforwardly defined as $\abs(x) = \ite{x>0}{x}{-x}$, we do not consider it in the overapproximations described in this paper.}, modulo.
We respectively denote these predicate and function symbols for integers by $\lessi, \lesseqi, \greateri, \greatereqi, \plusi, \minusi, \muli, \divi, \absi, \modi$ and for reals by $\lessr, \lesseqr, \greaterr, \greatereqr, \plusr, \minusr, \mulr, /_R$.
We use $\minusi$ and $\minusr$ to denote both the unary and binary minus.
Further, we use the same symbols $0, 1, -1,\dots$ for both integer and real constants.

In this paper we work with the quantifier-free fragment of first-order logic, denoted by the prefix ``QF\_". 
Free variables are implicitly assumed to be existentially quantified, hence the (SMT) variables correspond to first-order constants.
We denote the SMT variables by $a, b$; first-order variables by $x, y, z, w$; terms by $t, s, u, v$; formulas by $f$; all possibly with indices.
We use $t \lessi u \lesseqi v$ as an abbreviation for $t \lessi u \land u \lesseqi v$, and similarly with other comparison predicate symbols.
%
We use the standard notions of position in a formula and of (sub)formula polarity.
We denote the position in a formula by a sequence of positive integers.
We write $\epsilon$ for the empty sequence denoting the top-most position in a formula.
When $f'$ is a subformula of a formula $f$ at position $\pi$, where the top-level logical connective of $f'$ is $n$-ary, then the subformula of $f$ at position $\pi.i$ for any $i\in\{1,\dots,n\}$ is the $i$th argument of $f'$.
We denote the polarity of a (sub)formula of $f$ at position $\pi$ by $p_f(\pi)$.
Let $f$ be a formula and let $p_f(\epsilon)\bydef 1$.
Then, for each (sub)formula $f'$ at position $\pi$ of $f$:
If $f'$ has the form $f_1\land\dots\land f_n$ or $f_1\lor\dots\lor f_n$, then $\forall i \!\in\!\{1,\dots,n\}\!: p_f(\pi.i)\!=\!p_f(\pi)$. 
If $f'$ has the form $\lnot f_1$, then $p_f(\pi.1) = -p_f(\pi)$.
If $f'$ has the form $f_1\rightarrow f_2$, then $p_f(\pi.1) = -p_f(\pi)$ and $p_f(\pi.2) = p(\pi)$.
If $f'$ has the form $f_1\leftrightarrow f_2$, then $p_f(\pi.1)=p_f(\pi.2)=0$.
We say that the (sub)formula of $f$ at position $\pi$ has positive polarity if $p_f(\pi)=1$ and negative polarity if $p_f(\pi)=-1$.

We use the model domains of integers $\intg$ and real numbers $\real$, and we consider $\intg\subset\real$, i.e., we consider e.g.\ $1$ and $1.0$ to denote the same number.
We use $m, n$ to denote numbers from either $\intg$ or $\real$.
The theory predicate and function symbols are interpreted in the standard way, and we denote their interpretations by $<, \leq, >, \geq, +, -, *, div, abs, mod, /$, respectively.\footnote{
For positive $m$, $n\ div\ m$ is defined as $\lfloor n/m \rfloor$, for negative $m$ as $\lceil n/m \rceil$.}

\begin{example}\label{ex:1}
We illustrate the verification problem 
by a simplified Ethereum smart contract verification example.
Assume assets in the form of shares and money.
We model all the smart contract and specification variables, typically of type \texttt{uint256}, as non-negative integer variables.

Let there be a total supply of $t_S$ shares, altogether having the total monetary value of $t_M$, where $t_S\not=0$ and $t_M\not=0$.
If we withdrew $w_S$ shares, these shares would be sold and we would obtain the monetary value $w_M = (w_S*t_M)\ div\ t_S$.
Note that $div$ is the integer division: since the smart contract works with \texttt{uint256}s, we might get a little less money for the shares than the real value that they have.
Now, assume that $n_S$ new shares are added to the total supply $t_S$, 
having the same per-share value as the old shares (with precision up to the precision of $div$).
We obtain new total supply of shares $t_S^\prime = t_S + n_S$ and the new total monetary value $t_M^\prime = t_M + ((n_S*t_M)\ div\ t_S)$.
If we now withdrew the same number $w_S$ of shares from the new total supply, we would obtain the monetary value $w_M^\prime = (w_S* t_M^\prime)\ div\ t_S^\prime$.
Given this scenario, we would like to prove the following property: the monetary value from the original withdrawal is greater or equal to the monetary value from the 
second withdrawal,
i.e., $w_M \geq w_M^\prime$.


We encode this example in 
QF\_NIA.
We construct the verification condition for the example as a conjunction of all the assumptions and a negation of the conjecture:
\begin{equation}
    \begin{gathered}
    t_S\greateri0 \land t_M\greateri0 \land w_S\greatereqi 0\land w_M\greatereqi 0 \land n_S\greatereqi 0\land t_S^\prime\greatereqi 0\land t_M^\prime\greatereqi 0
    \land  w_M^\prime\greatereqi 0 \\
    \land\; w_M = (w_S\muli t_M)\ \divi\ t_S \; \land\;
    t_S^\prime = t_S \plusi n_S
    \land
    t_M^\prime = t_M \plusi ((n_S\muli t_M)\ \divi\ t_S) \\  \land\; 
    w_M^\prime = (w_S\muli t_M^\prime)\ \divi\ t_S^\prime
    \land
    \lnot( w_M \greatereqi w_M^\prime)
    \end{gathered}\label{eq:ex1}
\end{equation}

If the formula~\eqref{eq:ex1} is unsatisfiable, the property holds.
If~\eqref{eq:ex1} is satisfiable, its model constitutes a counterexample to the property.
In this case, the formula~\eqref{eq:ex1} is indeed unsatisfiable (it has no integer model), as will be later confirmed using the NRA overapproximation.
However, none of the SMT solvers we considered\footnote{\cvc5 v1.0.2~\cite{cvc5}, \mathsat v5.6.8~\cite{mathsat5}, \yices v2.6.4~\cite{yices}, and \zsol3 v4.11.0~\cite{Z3}}
can prove the unsatisfiability of its NIA encoding within a 5-minute time limit.
The SMT-LIB2 encoding of formula~\eqref{eq:ex1} is displayed in Appendix~\ref{sec:app1}.
\PH{Further, the smart contract verification problem this example is based on is also displayed in Appendix~\ref{sec:app3}.}
\end{example}

The Certora Prover uses formulas similar in character to~\eqref{eq:ex1}, and encodes them in the SMT-LIB2~\cite{SMTLIB} format.
\PH{
Wraparound semantics for \texttt{uint256}s are modelled by computing all operations that potentially under- or overflow modulo $2^{256}$.}
In addition to 
NIA, the Certora Prover also utilizes uninterpreted function symbols 
(UF) and the theory of datatypes (DT), resulting in the logic QF\_UFDTNIA.

\section{Overapproximation Methods}\label{sec:nra}
In this section, we describe the techniques for constructing the LIA and NRA overapproximations $\flia, \fnra$ of the formula $\fnia$ using QF\_NIA (and possibly other theories).
Formula $f$ is an \emph{overapproximation} of 
$\fnia$, 
if from unsatisfiability of $f$ follows that $\fnia$ is also unsatisfiable. 
We say that $f_2$ is a \emph{tighter} overapproximation of $\fnia$ than $f_1$ if both $f_1, f_2$ are overapproximations of $\fnia$, and the set of models of $f_1$ is a superset of models of $f_2$.
There is a trade-off between the overapproximation tightness and the practical complexity.
The overapproximation $f$ could precisely capture $\fnia$ 
-- but that would essentially force the solvers proving $f$ to reason with NIA instead of LIA or NRA.
Rather, we construct formulas $\flia, \fnra$ which can be solved using LIA, NRA, respectively, yet are typically unsatisfiable if $\fnia$ was unsatisfiable.
To construct an overapproximation of $\fnia$, we rewrite the original formula.
We denote rewriting of the term $t$ to the term $s$, i.e., replacing all occurrences of $t$ in a given formula by $s$, by \(t\rightsquigarrow s\).


\subsection{Linear Integer Arithmetic Overapproximation}
The LIA-based overapproximation proceeds in two steps.
First, it replaces nonlinear operations in $\fnia$ by replacing the NIA function symbols by uninterpreted function symbols as follows:
\[t \muli s \;\rightsquigarrow\; \umul(t, s) \qquad\quad
  t\ \divi\ s \;\rightsquigarrow\; \udiv(t, s) \qquad\quad
  t\ \modi\ s \;\rightsquigarrow\; \umod(t, s) 
  \]
The resulting formula $f_1$ is a very coarse overapproximation of $\fnia$.

Second, to tighten the overapproximation, the transformation adds axioms constraining $\umul,\udiv,\umod$ to the formula. 
To retain the overapproximation property, the axioms must correspond to valid properties of integer arithmetic -- e.g., an axiom regarding $\umul$ must only state valid properties of 
$\muli$. 
We do not add quantified axioms, but rather we instantiate axioms with selected terms occurring in $f_1$.
Due to space limitations we do not explicitly list all axioms, nor the instantiation strategy.
We rather list the axiomatized properties:
 (i) arithmetic axioms: commutativity, neutral and absorbing element with respect to $\umul$,  $\udiv$ and $\umod$, ranges of $\udiv$ and $\umod$ results;
 (ii) relating $\umul$ to the linear multiplication $\muli$ if one operand is an interpreted constant;
 (iii) overflow-related: multiplicative inverse property only holding if the multiplication did not overflow modulo $2^{256}$;\footnote{The formula $\fnia$ usually contains many occurrences of $t\ \modi\ 2^{256}$ (because $2^{256}-1$ is the maximal value of the integer type \texttt{uint256}), 
 and might also contain $0\lesseqi t$ and $t\lessi 2^{256}$ as the encoding of overflow checks.\label{ft:mod}}
 (iv) twos-complement related: certain standard identities regarding twos-complement multiplications in a modular ring;
 (v) relating pairs of multiplications: monotonicity and distributivity. 

These axioms are instantiatiated over single applications of the operators, as well as pairs of multiplications that lie on a common program path. They are not instantiated recursively, i.e., if an axiom produces a new multiplication, we don't generate an axiom for that.


Adding the axioms to $f_1$ results in a formula $\flia$ as a ground formula using the theories UF and LIA in addition to any theories that $\fnia$ used except for NIA.
E.g., if $\fnia$ is in QF\_DTNIA, $\flia$ will be in QF\_UFDTLIA.


\subsection{Non-Linear Real Arithmetic Overapproximation}
We construct the NRA overapproximation $\fnra$ of $\fnia$ in three steps.
The first two steps are similar to the LIA overapproximation: we replace the function symbols and add axioms.
Then we tighten the overapproximation by rewriting selected inequalities.
Similarly to $\flia$, 
the formula
$\fnra$ is ground and uses 
UF and NRA in addition to any theories that $\fnia$ used except for NIA.

In contrast to the LIA overapproximation, NRA overapproximation uses the real sort.
Hence, we change the sorts of all variables and functions from integers to reals.
We replace all interpreted integer constants by their real counterparts,
and the 
symbols $\plusi, \minusi, \muli$ by $\plusr, \minusr, \mulr$. 
We replace NIA function symbols $\divi, \modi$, 
which do not correspond to any interpreted function in NRA, using freshly added function symbols $\ufrac, \umod$, obtaining formula $f_1$: 
\begin{align*}
  t\ \divi\ s \;\rightsquigarrow\; t\divr s\ \minusr\ \ufrac(t, s) \qquad\qquad
  t\ \modi\ s \;\rightsquigarrow\; \umod(t, s) 
\end{align*}
Intuitively, we subtract $\ufrac(t,s)$ from $t\divr s$ to make sure that any integer model of $t\ \divi\ s$ will also be a model of its real translation.
Next we axiomatize the new function symbols.
We provide a definition for the function $\umod$, 
two axioms partially constraining $\ufrac$, and an axiom constraining all the real variables:
%
%
\begin{align}
    \umod(x, y) &\bydef {\small\begin{cases}
     x & \text{if }0\lesseqr x \lessr y \\
     x\minusr y & \text{if } y \lesseqr x \lessr 2\mulr y \\
     x\plusr y & \text{if } \minusr y \lesseqr x \lessr 0  \\
     y\mulr \ufrac(x, y) & \text{else}\end{cases}}
    \label{eq:defmod} \\
    \begin{split}
  \axfb(x, y) &\bydef (y \greaterr 0 \rightarrow 0 \lesseqr \ufrac(x, y) \lessr 1) \;\land\\
  &\qquad\quad(y \lessr 0 \rightarrow 0 \greatereqr \ufrac(x,y) \greaterr -1)
    \end{split}\label{eq:ax1} \\
  \axfz(x, y, z, w) &\bydef (x = z\!\mulr\!w \land (y\! =\! z\lor y\! =\! w)) \rightarrow \ufrac(x,y) = 0\label{eq:ax2} \\  
   \axia(x) &\bydef  x \lesseqr -1 \lor x=0 \lor x\greatereqr 1\label{eq:ax3}
\end{align}
The definition~\eqref{eq:defmod} is tailored to the occurrences of $\modi$ in $\fnia$ produced by the Certora Prover, as described in Footnote~\ref{ft:mod}.
%
%
Axioms~\eqref{eq:ax1},~\eqref{eq:ax2} constrain $\ufrac(x, y)$ to be between 0 and 1 (or 0 and $-1$ for negative $y$), and to be 0 when $y$ is known to be a divisor of $x$.
We instantiate them by selected terms occurring as arguments of $\divr,\umod$ and $\mulr$ in $f_1$.
%
Finally, we 
add an instance of~\eqref{eq:ax3} for each real variable in $f_1$,
to ensure that for all non-zero values $n, m$ both
$abs(n*m)\geq abs(n)$ and $abs(n*m)\geq abs(m)$ hold,
and for all $n$ and non-zero $m$ also $abs(n/m)\leq abs(n)$ holds, as it would for integer $n,m$.
%
We obtain the formula $f_2$ from $f_1$ by adding the definitions~\eqref{eq:defmod}-\eqref{eq:ax3}, and instances of axioms~\eqref{eq:ax1}-\eqref{eq:ax3} as described above.
\PH{Intuitively, $f_2$ is an overapproximation of $\fnia$, because any model of $\fnia$ can be extended to a model of $f_2$ by interpreting $\ufrac(n, m)$ as $n/m - (n\ div\ m)$ for integers $n, m$.
The instances of axioms~\eqref{eq:ax1}-\eqref{eq:ax3} hold in this model, because they are only instantiated for terms from $f_1$, which are in the considered model interpreted by integer values.}

The third step of the NRA overapproximation method is based on the following observation:
If we want to prove the conjecture $a \lesseqi b$ in NIA, we can do it by proving $a\minusr 1 \lessr b$ in NRA.
From the proof-by-refutation point of view, $\lnot (a\minusr 1\lessr b)$ is an overapproximation of $\lnot (a\lesseqi b)$.
%
Thus, to tighten the overapproximation $f_2$ of $\fnia$, we relax some of the inequalities in $f_1$ (not the inequalities in the axiom and function definitions~\eqref{eq:defmod}-\eqref{eq:ax3}).
%
We first replace each inequality $t \lessr s$ or $t \greaterr s$ occurring in the $f_1$ part of $f_2$ with positive polarity
by an equivalent inequality adding one more negation to obtain an inequality using $\greatereqr$ or $\lesseqr$ with negative polarity:
    \begin{align*}
      t \lessr s \;\rightsquigarrow\; \lnot(t\greatereqr s) \qquad\quad
      t \greaterr s \;\rightsquigarrow\; \lnot(t\lesseqr s) 
    \end{align*}
Then we relax each inequality $t\lesseqr s$ and $t\greatereqr s$ occurring in the $f_1$ part of $f_2$ with negative polarity, 
  obtaining the formula $\fnra$:
    \begin{align*}
      t \lesseqr s \;\rightsquigarrow\; t\minusr 1 \lessr s  \qquad\quad
      t \greatereqr s \;\rightsquigarrow\; t\plusr 1 \greaterr s 
    \end{align*}


\begin{example}\label{ex:4}
Using the method of this subsection on the formula~\eqref{eq:ex1} from Example~\ref{ex:1}, we obtain the formula consisting of the axiom definitions~\eqref{eq:ax1}-\eqref{eq:ax3} (since~\eqref{eq:ex1} does not use $\umod$, 
we omit the definition~\eqref{eq:defmod}) 
and the following:
{\small
\begin{equation*}
    \begin{gathered}
    \lnot(t_S\!\minusr\! 1\lessr0) \land \lnot(t_M\!\minusr\! 1\lessr0) \land w_S\greatereqr 0\land w_M\greatereqi 0\land n_S\greatereqr 0\land t_S^\prime\greatereqr 0
    \land t_M^\prime\greatereqr 0 \land  w_M^\prime\greatereqr 0 \\
    \land\, w_M\!=\!(w_S\!\mulr\!t_M)\divr t_S \minusr \ufrac(w_S\!\mulr\!t_M, t_S)
    \land
    t_M^\prime \!=\! t_M \plusr ((n_S\!\mulr\!t_M)\divr t_S \minusr \ufrac(n_S\!\mulr\!t_M, t_S)) \\
    \land\; t_S^\prime = t_S \plusr n_S \,\land\,
    w_M^\prime = (w_S\mulr t_M^\prime)\divr t_S^\prime \ \minusr\ \ufrac(w_S\mulr t_M^\prime, t_S^\prime) \,
    \land\,
    \lnot( w_M\plusr 1\greaterr w_M^\prime) \\
    \land \axfb(w_S\!\mulr\!t_M, t_S)
    \!\land\! \axfz(w_S\!\mulr\!t_M, t_S, w_S, t_M)
    \!\land\! \axfz(w_S\!\mulr\!t_M, t_S, n_S, t_M) \\
    \land\axfz(w_S\!\mulr\!t_M, t_S, w_S, t_M^\prime)
    \!\land\! \axfb(n_S\!\mulr\! t_M, t_S)
    \!\land\! \axfz(n_S\!\mulr\! t_M, t_S, w_S, t_M) \\
    \land\axfz(n_S\!\mulr\! t_M, t_S, n_S,t_M)
    \!\land\! \axfz(n_S\!\mulr\! t_M, t_S, w_S,t_M^\prime)
    \!\land\! \axfb(w_S\!\mulr\! t_M^\prime, t_S^\prime) \\
    \land\, \axfz(w_S\!\mulr\! t_M^\prime, t_S^\prime, w_S,t_M)
    \land \axfz(w_S\!\mulr\! t_M^\prime, t_S^\prime, n_S,t_M) \land \axia(t_S) \\
    \land \axfz(w_S\!\mulr\! t_M^\prime, t_S^\prime, w_S,t_M^\prime)
    \!\land\!\axia(t_M)\!\land\!\axia(w_S) \!\land\! \axia(w_M)\\
     \land \axia(n_S)\!\land\!\axia(t_S^\prime) \!\land\!\axia(t_M^\prime)\!\land\!\axia(w_M^\prime)
    \end{gathered}\label{eq:ex4}
\end{equation*}}%
When encoded in the SMT-LIB2 syntax, this formula is proved unsatisfiable by the SMT solvers \yices and \cvc5.
Its SMT-LIB2 encoding is displayed in Appendix~\ref{sec:app2}.
\end{example}

\section{Experimental Evaluation}\label{sec:experiments}

\paragraph{Experimental setup.}
Our experimental setup mirrors the setup used by the Certora Prover.
We considered the NIA encoding of 792 verification problems stemming from customer contracts that were under verification by Certora, and compared the performance of four SMT solvers on these benchmarks against overapproximated versions of the benchmark set.

The LIA overapproximation method was implemented directly in the Certora Prover.
The Certora Prover hence produces both the natural NIA encoding of the problems and their LIA overapproximations.
In order to prototype the NRA overapproximation, we implemented the method as a formula-to-formula transformation in Python using the tokenizer of \solver{PySMT}~\cite{pysmt2015}.
The formulas produced by the Certora Prover and our artifact are in SMT-LIB2 format~\cite{SMTLIB-standard}.
All benchmarks are available at \url{https://github.com/Certora/PublicBenchmarks/tree/master/Sep2022}.
The artifact for NRA overapproximation is available at \url{https://github.com/hzzv/scripts/tree/main/real_relaxation}.

The 792 benchmarks originate from two different encodings of 396 problems: one encoding in QF\_UFNIA, one in QF\_UFDTNIA.
The encodings differ in how they model hash functions used by smart contracts for memory addressing.
The QF\_UFNIA encoding represents the hash functions using uninterpreted functions while the QF\_UFDTNIA encoding uses inductive datatypes.
Since the hash functions are only used for memory layout, the hash function representation mostly does not interact with non-linear arithmetic.
Thus, there is in general no interplay of datatypes and NIA in the QF\_UFDTNIA benchmark set.
Out of both benchmark sets, we excluded 300 benchmarks found to be satisfiable in QF\_UFNIA or QF\_UFDTNIA during our experiments, resulting in 246 benchmarks in each set.
For these benchmarks we generated the overapproximation sets QF\_UFLIA, QF\_UFDTLIA, and QF\_UFNRA, QF\_UFDTNRA.
In the sequel we refer to the sets QF\_UFNIA, QF\_UFLIA, QF\_UFNRA together as QF\_UF+, and to the sets QF\_UFDTNIA, QF\_UFDTLIA, QF\_UFDTNRA together as QF\_UFDT+.

We compared the performance of 
\cvc5 v1.0.2~\cite{cvc5}, \mathsat v5.6.8~\cite{mathsat5}, \yices v2.6.4~\cite{yices}, and \zsol3 v4.11.0~\cite{Z3} on the original and the overapproximated benchmarks.
These are all the non-portfolio and non-wrapper SMT solvers that competed in SMT-COMP 2022~\cite{SMT-COMP} in the single query track of division QF\_NonLinearIntArith. 
We ran all the experiments on computers with 32 cores (AMD Epyc 7502, 2.5 GHz) and 1 TB RAM using the benchmarking tool \solver{BenchExec}~\cite{benchexec}.
We used the time limit of 300 seconds and the memory limit of 16 GB per problem, corresponding to the time limits used in practice by the Certora Prover.

\paragraph{Experimental results.}


From the 492 benchmarks, 482 use integer division and 490 use modulo.
Multiplication is used on average 83 times per problem, 
division 37 times per problem, 
and modulo 47 times per problem. 
Both overapproximations therefore added many instances of axioms related to multiplication, division and modulo.
Further, 
since the Certora Prover internally converts the smart contracts to a static-single-assignment form and hence introduces many auxiliary variables 
(the average number of variables in a problem is 1361), 
the NRA overapproximation also added many instances of the axiom~\eqref{eq:ax3}.
Consequently, the LIA and NRA benchmarks were on average 70\% and 243\% larger in file size, respectively, than the original benchmarks.
We note that even though instantiating the axioms increases the benchmark size significantly,
in our experience it leads to better solver performance than using quantified axioms instead. 

We say that a benchmark was solved to convey that it was proved unsatisfiable.
The main goal of using the overapproximations is to verify smart contracts that were not verified in the NIA encoding.
The key metric is therefore the number of \emph{newly solved problems},
i.e., problems solved in an overapproximation but not solved in the NIA encoding, no matter which solver solves them.
Hence in the result analysis, we use the concept of a \emph{virtual best solver} (VBS) -- a hypothetical solver which solves each problem as fast as the fastest solver we ran.
E.g., if we have two benchmarks, one benchmark solved by \cvc5 in 1 second and by \zsol3 in 2 seconds, and another benchmark solved only by \zsol3 in 1 second, then the VBS is considered to have solved both benchmarks with the runtime of 1 second for each of them.

\begin{figure}[t]
\begin{minipage}{0.54\textwidth}
\renewcommand{\arraystretch}{1.1} 
    \centering
{\small
        \begin{tabular}{@{\extracolsep{5pt}}r>{\centering}p{0.5cm}>{\centering}p{1.35cm}>{\centering\arraybackslash}p{1.35cm}}
        QF\_UF+   & NIA       & LIA\,(new)& NRA\,(new) \\ \hline
        \cvc5       & 8         & 7 (7)     & \bf 13 (13)  \\
        \mathsat    & 10        & 1 (1)     & \bf 40 (31) \\
        \yices      & 0         & 7 (7)     & \bf 34 (34)  \\
        \zsol3      & \bf 47    & 25 (9)       & 20 (7)    \\ \hline
        virtual best solver & \bf 52 & 34 (11) & 42 (14)  \\ \hline\hline
        QF\_UFDT+   & NIA       & LIA\,(new)& NRA\,(new) \\ \hline
        \cvc5       & 9     & 8 (6) & \bf 43 (39) \\
        \zsol3      & \bf 55 & 49 (19) & 25 (13) \\ \hline
        virtual best solver & \bf 57 & 52 (19) & 47 (16) \\
    \end{tabular}}
    \vspace*{0.5em}
    \caption{Numbers of solved problems in the respective benchmark sets.
    Columns ``LIA" and ``NRA" refer to the benchmark overapproximations.
    In the parentheses is the number of problems solved in the respective overapproximated benchmark set, but not in the original NIA benchmark set.
    The ``virtual best solver" row corresponds to the best result for each benchmark -- i.e., the numbers in this row are the counts of benchmarks solved by any solver.
    }
    \label{tab:cresults}
\end{minipage}
\begin{minipage}{0.01\textwidth}\phantom{.}\end{minipage}
\begin{minipage}{0.01\textwidth}\phantom{.}\end{minipage}
\begin{minipage}{0.43\textwidth}
\centering
    \includegraphics[scale=0.5]{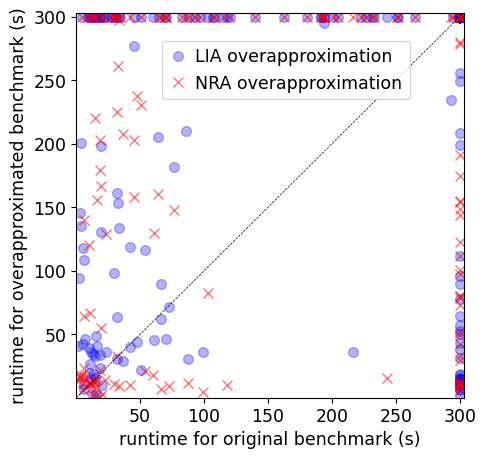}
    \caption{Comparison of the runtimes on original and overapproximated benchmark sets.
    We compare runtimes for the virtual best solver -- for each problem we consider the best runtime achieved by any of the four solvers.
      }
    \label{fig:cplots}
\end{minipage}
\end{figure}

We ran all four SMT solvers on the benchmarks in QF\_UF+, but since \mathsat and \yices do not support datatypes, only the solvers \cvc5 and \zsol3 on the benchmarks in\linebreak QF\_UFDT+.
The results are displayed in Figure~\ref{tab:cresults}.
Both the LIA and NIA overapproximations led to 30 newly solved benchmarks compared to the original benchmark set (28\% increase).
If we combined the results for both overapproximations, the VBS solved 50 and 59 benchmarks in at least one of the overapproximations originating from QF\_UFNIA and QF\_UFDTNIA, respectively.
19 and 23 benchmarks were newly solved in at least one of the overapproximations of QF\_UFNIA and QF\_UFDTNIA, respectively, corresponding to a 39\% increase in the number of smart contracts we were able to verify compared to the NIA encoding.

Interestingly, the NRA overapproximation significantly increased the number of solved benchmarks for all solvers except for \zsol3.
A large majority of these benchmarks were newly solved, suggesting that the techniques used by the solvers for 
NIA are orthogonal to our NRA overapproximation, and that the solvers might potentially benefit from using a similar method.

Figure~\ref{fig:cplots} displays a comparison of the runtimes for the VBS for all the problems in the original and the two overapproximated benchmark sets.
We used the actual best runtime for problems solved as unsatisfiable, but for problems with only non-unsatisfiable results we set the runtime to 300 seconds to convey that none of the solvers was able to solve the problem in the given time limit.
There were 56 problems solved as unsatisfiable in both the original and the LIA overapproximated set, and 59 such problems for the original and the NRA overapproximated set.
40 problems were solved faster in the original set compared to LIA overapproximations, and 35 faster in the original set compared to NRA overapproximations.

\paragraph{Measuring the Overapproximation Tightness.}\label{sec:measure}
In order to measure how tight are both overapproximations, we look at the numbers of problems 
solved as unsatisfiable in the original set, but either solved 
as satisfiable or not solved at all in the overapproximated set.
Out of the 109 originally unsatisfiable problems, in the LIA overapproximation no problems were solved as satisfiable, while 53 were not solved at all.
In the NRA overapproximation, 24 originally unsatisfiable problems were solved as satisfiable and 26 were not solved at all.
We thus conclude that the LIA overapproximation is quite tight and is unlikely to benefit from adding more axioms.
On the other hand, the NRA overapproximation could be tightened.\footnote{
We tried using a modification of the axiom~\eqref{eq:ax1} bounding $\ufrac(x, y)$ to be at most $1-_R 1/_Ry$ for $y>_R0$ and at least $-1-_R 1/_Ry$ for $y<_R0$. However, compared to~\eqref{eq:ax1}, this tightening did not increase the number of newly solved problems, nor decrease the number of originally unsatisfiable problems being solved as satisfiable.}

\PH{
\paragraph{Applications in Other Domains.}
We also tried running our NRA overapproximation tool on the QF\_NIA and QF\_UFNIA sets of the SMT-LIB benchmark library~\cite{SMTLIB}.
However, the QF\_NIA set only contains a small number of problems using either division or modulo, and the QF\_UFNIA set only contains non-linear operations in function definitions, for which the method does not add any axiom instances.
Hence, it is not surprising that the SMT solvers were in general less successful on the overapproximated benchmarks compared to the original benchmarks.
The only notable exception was the performance of Yices on the QF\_NIA benchmarks, where out of 10343 total problems it solved 5243 in the original set, but 5992 benchmarks in the overapproximated set.
Therefore, we conclude that for successful application of overapproximation methods, it is crucial to fine-tune the methods for the specific domain.
}

\section{Related Work}\label{sec:related}
Contemporary SMT solvers implement a range of methods for reasoning with QF\_NIA, 
such as
bit-blasting~\cite{fuhs2007} (\solver{AProVe}~\cite{fuhs2007}, \cvc5~\cite{cvc5}, \zsol3~\cite{Z3}, and \solver{SMT-RAT}~\cite{SMT-RAT}), linearization~\cite{borralleras2012sat} (\solver{Barcelogic}~\cite{bofill2008barcelogic}), incremental linearization~\cite{cimatti2018} (\mathsat~\cite{mathsat5}, \cvc5, and \zsol3), and NRA overapproximation combined with branch-and-bound~\cite{kremer2016generalised,jovanovic2017solving} (\solver{SMT-RAT}, \zsol3, and \yices~\cite{yices}).

The overapproximations we evaluate in this case study are similar in spirit to that of~\cite{cimatti2018,kremer2016generalised,jovanovic2017solving} (not~\cite{borralleras2012sat}, since that is geared towards finding models, while the overapproximations we focused on aim for preservation of unsatisfiability).
However, all these approaches use iterative refinements of the overapproximation, while the overapproximation methods we compared work on top of SMT solvers: they only create one overapproximation and pass it to a solver supporting LIA or NRA.
Further, the overapproximation methods we focused on are specialized for the domain of smart contract verification  (e.g., using a specialized definition of $\umod$).

\section{Conclusions}
In this case study we focused on overapproximation methods for verification of smart contracts using NIA.
We described and evaluated two overapproximation methods using LIA and NRA.
Our results show that both methods lead to solving a large number of industry benchmarks that were not solved in their natural NIA encoding, emphasizing the benefit of domain-specific verification methods.
Further, our evaluation indicates that the NRA overapproximation can be further refined.
Finally, our results also suggest that the NRA overapproximation method could be combined with existing methods for real overapproximation used by SMT solvers.


\paragraph{Acknowledgements.}
We thank J\'an Hozza, Laura Kov\'acs, and Gereon Kremer for fruitful discussions.
We also thank Shelly Grossman, Jochen Hoenicke, and Mooly Sagiv.
This work was partially funded by the ERC CoG ARTIST 101002685, and the FWF grants LogiCS W1255-N23 and LOCOTES P 35787.
%
%
%
\bibliographystyle{splncs04}
\bibliography{bibliography}

\newpage
\appendix

\section{Appendix}\label{sec:appendix}
\subsection{SMT-LIB Encoding of Example~\ref{ex:1}}\label{sec:app1}
\setlength{\columnseprule}{0.5pt}
To encode our running example into SMT-LIB2, we rename the variables $t_S, t_S^\prime$, $t_M, t_M^\prime$, $w_M, w_M^\prime$, $n_S, w_S$ to \texttt{ts1}, \texttt{ts2}, \texttt{tm1}, \texttt{tm2}, \texttt{wm1}, \texttt{wm2}, \texttt{ns}, \texttt{ws}, respectively.

The encoding of the formula~\eqref{eq:ex1} from Example~\ref{ex:1}:
\begin{multicols}{2}
\begin{lstlisting}[basicstyle=\ttfamily\footnotesize]
(set-logic QF_NIA)

(declare-const ts1 Int)
(declare-const ts2 Int)
(declare-const tm1 Int)
(declare-const tm2 Int)
(declare-const wm1 Int)
(declare-const wm2 Int)
(declare-const ns Int)
(declare-const ws Int)

(assert (and
  (> ts1 0)
  (> tm1 0)
  (>= ws 0)
  (>= wm1 0)
  (>= ns 0)
  (>= ts2 0)
  (>= tm2 0)
  (>= wm2 0)
))

(assert (= wm1 (div (* ws tm1) ts1)))
(assert (= ts2 (+ ts1 ns)))
(assert
  (= tm2 (+ tm1 (div (* ns tm1) ts1))))
(assert (= wm2 (div (* ws tm2) ts2)))
(assert (not (>= wm1 wm2)))

(check-sat)
\end{lstlisting}
\end{multicols}

\subsection{SMT-LIB Encoding of Example~\ref{ex:4}}\label{sec:app2}
We overapproximated the SMT-LIB2 formula above using our script.
The script considers relaxing each inequality individually, resulting in the inequalities $t_S-_R1\geq 0$ and $t_M-_R1\geq 0$ instead of the equivalent $\lnot(t_S\!-_R\! 1<_R0)$ and $\lnot(t_M\!-_R\! 1<_R0)$ from formula~\eqref{eq:ex4}.
We also added comments annotating parts of the encoding.
Otherwise the resulting formula matches the complete relaxed formula from Example~\ref{ex:4}:

\begin{multicols}{2}
\begin{lstlisting}[basicstyle=\ttfamily\footnotesize]
(set-logic QF_UFNRA)

;; Newly added functions
(declare-fun ufrac (Real Real) Real)
(define-fun ax_frac_bound
  ((x Real) (y Real)) Bool
  (and
    (=> (> y 0) (and
        (<= 0.0 (ufrac x y))
        (< (ufrac x y) 1.0)))
    (=> (< y 0) (and
        (>= 0.0 (ufrac x y))
        (> (ufrac x y) (- 1.0))))
))
(define-fun ax_frac_zero
  ((x Real) (y Real) (z Real) (w Real))
  Bool
  (=> (and (= x (* z w))
           (or (= y z) (= y w)))
      (= (ufrac x y) 0))
)
(define-fun ax_int_approx 
  ((x Real)) Bool
  (or (= x 0) (>= x 1) (<= x (- 1)))
)

;; Translated declarations
(declare-const ts1 Real)
(declare-const ts2 Real)
(declare-const tm1 Real)
(declare-const tm2 Real)
(declare-const wm1 Real)
(declare-const wm2 Real)
(declare-const ns Real)
(declare-const ws Real)

;; Newly added axiom instances:
(assert (ax_frac_bound (* ws tm2) ts2))
(assert
  (ax_frac_zero (* ws tm2) ts2 ws tm2))
(assert
  (ax_frac_zero (* ws tm2) ts2 ns tm1))
(assert
  (ax_frac_zero (* ws tm2) ts2 ws tm1))
(assert (ax_frac_bound (* ns tm1) ts1))
(assert
  (ax_frac_zero (* ns tm1) ts1 ws tm2))
(assert
  (ax_frac_zero (* ns tm1) ts1 ns tm1))
(assert
  (ax_frac_zero (* ns tm1) ts1 ws tm1))
(assert (ax_frac_bound (* ws tm1) ts1))
(assert
  (ax_frac_zero (* ws tm1) ts1 ws tm2))
(assert
  (ax_frac_zero (* ws tm1) ts1 ns tm1))
(assert
  (ax_frac_zero (* ws tm1) ts1 ws tm1))
(assert (ax_int_approx wm2))
(assert (ax_int_approx ts1))
(assert (ax_int_approx ns))
(assert (ax_int_approx ws))
(assert (ax_int_approx wm1))
(assert (ax_int_approx tm1))
(assert (ax_int_approx tm2))
(assert (ax_int_approx ts2))

;; Relaxed assertions:
(assert (and
  (>= (- ts1 1) 0)
  (>= (- tm1 1) 0)
  (>= ws 0)
  (>= wm1 0)
  (>= ns 0)
  (>= ts2 0)
  (>= tm2 0)
  (>= wm2 0)
))

(assert (= wm1
  (- (/ (* ws tm1) ts1)
     (ufrac (* ws tm1) ts1))))
(assert (= ts2 (+ ts1 ns)))
(assert (= tm2
  (+ tm1 (- (/ (* ns tm1) ts1)
            (ufrac (* ns tm1) ts1)))))
(assert (= wm2
  (- (/ (* ws tm2) ts2)
     (ufrac (* ws tm2) ts2))))
(assert (not (> (+ wm1 1) wm2)))

(check-sat)
\end{lstlisting}
\end{multicols}

\subsection{A Sample Smart Contract Verification Problem}\label{sec:app3}
Example~\ref{ex:1} corresponds to a part of a simplified version of the following smart contract verification problem.
The variables $t_S, t_M, w_M, w_M^\prime, n_S, w_S$ correspond to variables
\texttt{elastic1}, \texttt{base1}, \texttt{toBase1}, \texttt{toBase2}, \texttt{addAmount}, \texttt{someValue}, respectively.

The original specification in the Certora Verification Language syntax:\footnote{\url{https://docs.certora.com/en/latest/docs/cvl/index.html}}
\begin{multicols}{2}
\begin{lstlisting}[basicstyle=\ttfamily\footnotesize]
methods {
  function getElastic() external
    returns (uint128) envfree;
  function getBase() external
    returns (uint128) envfree;
  function toBaseFloor(uint256 elastic)
    external returns (uint256) envfree;
  function addFloor(uint256 elastic)
    external returns (uint256) envfree;
}

/** x and y are almost equal; y may be
 *  smaller than x up to epsilon */
definition only_slightly_larger_than(
    uint x, uint y, uint epsilon)
    returns bool = 
  y <= x &&
    x <= require_uint256(y + epsilon);

/** Check that adding some amount does
 *  not impact the elastic/base
 *  quotient beyond the given error
 *  margin. */
rule integrityOnAdd {
  uint128 addAmount;
  uint128 someValue;

  uint128 elastic1 = getElastic();
  uint128 base1 = getBase(); 
  /** these cases are handled
   *  separately */
  require base1 != 0;
  require elastic1 != 0;

  uint256 toBase1 =
    toBaseFloor(someValue);

  /** using ~Floor version so the
   *  rounding error is in favour
   *  of the "bank" */
  uint256 addAmountBase =
    addFloor(addAmount);

  uint256 toBase2 =
    toBaseFloor(someValue);

  uint256 sum1 = require_uint256(
    elastic1 + addAmount);
  uint256 error_margin =
    require_uint256(
      (someValue / sum1) + 1);

  assert only_slightly_larger_than(
    toBase1, toBase2, error_margin), 
      "addFloor(..) may not change the
       result of toBaseFloor(someValue)
       significantly; but does so.";
}
\end{lstlisting}
\end{multicols}

The corresponding contract in Solidity:
\begin{multicols}{2}
\begin{lstlisting}[basicstyle=\ttfamily\footnotesize]
pragma solidity 0.6.12;
pragma experimental ABIEncoderV2;

/** verification harness contract */
contract RebaseWrapper {
  using BoringMath for uint256;
  using BoringMath128 for uint128;

  Rebase public rebase;

  function getElastic() public view
      returns (uint128) {
    return rebase.elastic;
  }

  function getBase() public view
      returns (uint128) {
    return rebase.base;
  }

  function toBaseFloor(uint256 elastic)
      public view returns
      (uint256 base) {
    if (rebase.elastic == 0) {
      base = elastic;
    } else {
      base = elastic.mul(rebase.base) /
             rebase.elastic;
    }
  }

  function addFloor(uint256 elastic)
    public returns (uint256 base) {
    base = toBaseFloor(elastic);
    rebase.elastic =
      rebase.elastic.add(
        elastic.to128());
    rebase.base =
      rebase.base.add(base.to128());
    return base;
  }

}

/** from BoringMath.sol */

/** @notice A library for performing
 *  over-/underflow-safe math, updated
 *  with awesomeness from DappHub
 *  https://github.com/dapphub/ds-math
 */
library BoringMath {
  function mul(uint256 a, uint256 b)
      internal pure returns
      (uint256 c) {
    require(b == 0 ||
            (c = a * b) / b == a,
      "BoringMath: Mul Overflow");
  }

  function to128(uint256 a) internal
      pure returns (uint128 c) {
    require(a <= uint128(-1),
      "BoringMath: uint128 Overflow");
    c = uint128(a);
  }
}

/** @notice A library for performing
 *  over-/underflow-safe addition
 *  and subtraction on uint128. */
library BoringMath128 {
  function add(uint128 a, uint128 b)
      internal pure returns
      (uint128 c) {
    require((c = a + b) >= b,
      "BoringMath: Add Overflow");
  }
}

/** from BoringRebase.sol */
struct Rebase {
  uint128 elastic;
  uint128 base;
}
\end{lstlisting}
\end{multicols}
\noindent Note that the code and comments above were edited for conciseness and clarity.
The code includes the smart contract written by Certora during the verification process, and the relevant parts from libraries \texttt{BoringMath.sol} and \texttt{BoringRebase.sol}, version corresponding to commit \texttt{9f6d870}\footnote{\url{https://github.com/boringcrypto/BoringSolidity/tree/9f6d8708aa5df8b5d6cdad4a917e37b14c348684/contracts/libraries}}.




\end{document}